\documentclass[%
reprint,
longbibliography,
eprint,
superscriptaddress,
a4paper,
twoside,
aps,
pre,
showkeys,
floatfix,
]{revtex4-2}

\usepackage[utf8]{inputenc}

\usepackage{amsmath,amssymb,amsfonts}
 \usepackage{mathtools}
\usepackage{bm}
\usepackage{graphicx}

\usepackage{url}
\usepackage{hyperref}
\usepackage{xcolor}

\newcommand\rme{\ensuremath{\mathrm e}}
\newcommand\rmd{\ensuremath{\mathrm d}}

\newcommand\noiseStrength{\ensuremath\xi}
\newcommand\xAv{\ensuremath{\overline{x}}}
\newcommand\vAv{\ensuremath{\overline{v}}}
\newcommand\zAv{\ensuremath{\overline{z}}}
\newcommand\connectivity{\ensuremath{\lambda_d}}

\newcommand\wI{\ensuremath{w(0)}}
\newcommand\xI{\ensuremath{x(0)}}
\newcommand\xIsq{\ensuremath{x^2(0)}}

\newcommand\mean[1]{\ensuremath{\left\langle #1 \right\rangle}}

\newcommand\Sect[1]{Sect.~\ref{#1}}

\newcommand\Eq[1]{Eq.~\eqref{eq:#1}}
\newcommand\Eqs[1]{Eqs.~\eqref{eq:#1}}

\newcommand\Fig[1]{Fig.~\ref{fig:#1}}

\begin{document}


\title[Complex networks of interacting stochastic tipping elements]
{Complex networks of interacting stochastic tipping elements: \\
cooperativity of phase separation in the large-system limit}

\author{Jan Kohler}
\affiliation{\text{Institute for Theoretical Physics, University of Leipzig, 04103 Leipzig, Germany, EU}}
\affiliation{\text{Earth System Analysis, Potsdam-Institute for Climate Impact Research, Member of the Leibniz Association,}
  \text{14473 Potsdam, Germany, EU}}
  
\author{Nico Wunderling}
\email[correspondence should be addressed to:\\]{wunderling@pik-potsdam.de, juergen.vollmer@uni-leipzig.de}
\affiliation{\text{Earth System Analysis, Potsdam-Institute for Climate Impact Research, Member of the Leibniz Association,}
  \text{14473 Potsdam, Germany, EU}}
\affiliation{\text{Institute of Physics and Astronomy, University of Potsdam, 14476 Potsdam, Germany, EU}}
\affiliation{\text{Department of Physics, Humboldt University of Berlin, 12489 Berlin, Germany, EU}}

\author{Jonathan F. Donges}
\affiliation{\text{Earth System Analysis, Potsdam-Institute for Climate Impact Research, Member of the Leibniz Association,}
  \text{14473 Potsdam, Germany, EU}}
\affiliation{\text{Stockholm Resilience Centre, Stockholm University, 10691 Stockholm, Sweden, EU}}

\author{Jürgen Vollmer}
\email[correspondence should be addressed to:\\]{wunderling@pik-potsdam.de, juergen.vollmer@uni-leipzig.de}
\affiliation{\text{Institute for Theoretical Physics, University of Leipzig, 04103 Leipzig, Germany, EU}}

\date{\today}

\begin{abstract}
  
  Tipping elements in the Earth System receive increased scientific attention over the recent years due to their nonlinear behavior and the risks of abrupt state changes. While being stable over a large range of parameters, a tipping element undergoes a drastic shift in its state upon an additional small parameter change when close to its tipping point. Recently, the focus of research broadened towards emergent behavior in networks of tipping elements, like global tipping cascades triggered by local perturbations. Here, we analyze the response to the perturbation of a single node in a system that initially resides in an unstable equilibrium. The evolution is described in terms of coupled nonlinear equations for the cumulants of the distribution of the elements. We show that drift terms acting on individual elements and offsets in the coupling strength are sub-dominant in the limit of large networks, and we derive an analytical prediction for the evolution of the expectation (i.e., the first cumulant). It behaves like a single aggregated tipping element characterized by a dimensionless parameter that accounts for the network size, its overall connectivity, and the average coupling strength. The resulting predictions are in excellent agreement with numerical data for Erdös-Rényi, Barabási-Albert and Watts-Strogatz networks of different size and with different coupling parameters.
  
\end{abstract} 


\maketitle

\section{Introduction}

Hysteresis is a hallmark of first-order phase transitions.
For thermodynamic systems it leads to supercooling and superheating with subsequent explosive phase changes~\citep{safari2017review,wunderlich2007one}.
These rapid changes are well-understood for common thermodynamic phase transitions~\citep[e.g.][]{binder1987theory,kuwahara1995first,onuki2002book}.
However, they still pose challenges for systems with non-standard interaction rules, like the Achlioptas Process~\citep{daCosta2010,DSouza2010,DSouza2015,Grassberger2011,Riordan2011}, or non-standard interaction topologies like processes on networks, where hysteresis can lead to cascading failure~\citep{Buldyrev2010,Motter_Timme2018,Watts2002,abraham1991computational}.
Hysteresis and explosive transitions between (meta-)stable states
are also commonly observed in other systems.
In ecological and climate systems~\cite{Lenton2019} 
and in finance, economics and politics~\citep{Brummitt2015}
they
are commonly denoted as tipping processes. 
Further applications are discussed in 
the recent 
reviews
\citep{FeudelReview2018,Motter_Timme2018}. 

The present study is inspired by current models of climate change
that are formulated in terms of networks of
interacting tipping elements~\citep{Scheffer1989,
  Hughes2013, Lenton2013,Rocha2018,Zemp2017,
  Steffen2018, Wunderling2020, Kroenke2020}.
The interactions provide long-lived metastable states and rapid cooperative transitions between the states.
Often such a stabilization and cooperation in a tipping event is caused by positive feedback effects,
i.e., coupling or interactions
that tend to align individual elements.
For ferromagnetic systems the interaction tends to align individual spins with the molecular field. 
An example in the Earth System is the surface albedo of sea ice~\citep{garbe2020hysteresis,Robinson2012}.
A decrease in the ice-covered surface area due to an increase in global mean temperature decreases the surface albedo.
This, in turn, increases the temperature and causes higher rates of melting~\citep{Curry1995,Wackerbauer2011}.
The interaction between the elements leads to \textit{cascading} behavior when the abrupt state shift of an element causes the tipping of another~\citep{Kriegler2009}.
The resulting rapid nonlinear changes of the climate have been predicted almost 40 years ago~\citep{Watson1983}.
Early on they were investigated regarding a Snowball-Earth/warm-Earth-state transition~\citep{budyko1969effect,sellers1969global,hoffman1998neoproterozoic,lucarini2019transitions},
and more recently towards a potential \textit{hothouse} state~\citep{Steffen2018,Lenton2019}.

\begin{figure*}
  \centering
  \includegraphics[width=0.9\textwidth]{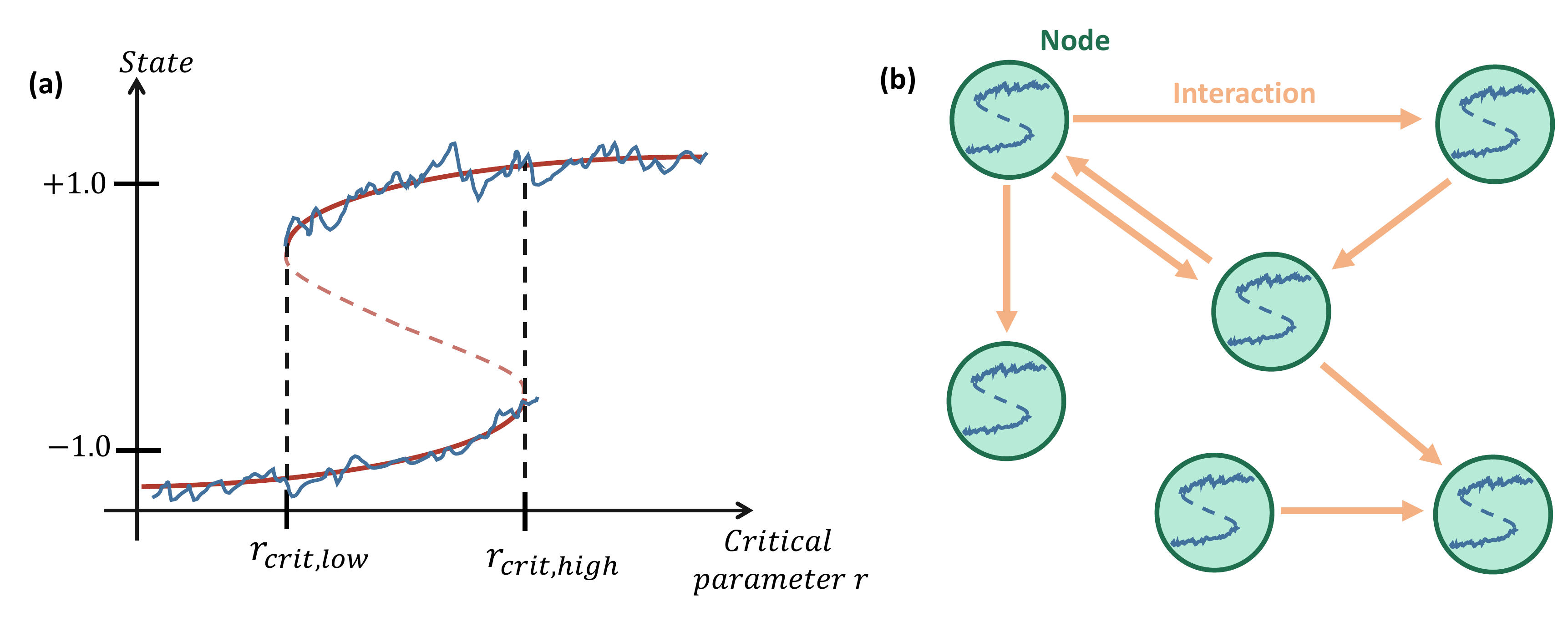}
  \caption{
    \textit{Dynamics of the nodes in a network of tipping elements.}
    (a) Bifurcation diagram of Eq.~\eqref{eq:normal_single_node}
    where solid and dashed red lines show the positions of the stable and unstable steady states, respectively.
    In the range between
  $r_\text{crit,\ low}$ and $r_\text{crit,\ high}$
  there are two stable and one unstable state.
    At the borders of the interval 
    a stable state merges with the unstable state, and they disappear in a saddle-node bifurcation,
  and systems at that position will decay to the remaining stable state.
    The trembling blue line illustrates the stochastic fluctuations around the stable state that we are using here, see Eq.~\eqref{eq:system}.
    (b) Nodes in an exemplary 
    network of 
    interacting tipping elements (see Eq.~\eqref{eq:system}), where each node is of the form shown in panel (a)).}
  \label{fig:model}
\end{figure*}

The present paper adds to the ongoing research on the dynamic behavior of coupled tipping elements by identifying universal behavior in the tipping of large-scale systems.
Tipping in a network comprising a large number of tipping elements
will be characterized by the first two cumulants (expectation 
and variance) of the distribution of tipping elements.
In large networks 
the expectation 
behaves like a single aggregated tipping element.
Moreover, 
  the variance remains small,
  and for small noise it rapidly decays
except for an initial transient growth early on.
Hence, we adopted the expectation as order parameter of the transition,
  and we will show that the tipping dynamics depends only on two dimensionless control parameters.
These parameters take an analogous rise to the temperature and the external field
  in the thermodynamic treatment of magnetic 
  phase transitions \citep{Callen1985Book}.

The paper is organized as follows.
In \Sect{sec:Theory} we introduce the system of nonlinear differential equations
that are analyzed in the present study,
and we describe how they are coupled in network structures.
Moreover, we also describe how the systems are simulated. 
In \Sect{sec:evolution} we present our main results.
First, the cumulants of the distribution of states of a network of unbiased tipping elements are defined,
and their time evolution is derived (\Sect{subsec:moment-expansion}). 
Then, we introduce rescaled coordinates
that allow us to discuss universal aspects of the late-time dynamics (\Sect{subsec:order-parameter-asymptotics}).
In \Sect{sec:late-time-evolution} the dynamics of the
expectation 
is worked out and compared to numerical data.
\Sect{sec:drift-impact} addresses the impact of additional forcing terms (bias) and offsets in the coupling coefficients. 
In \Sect{sec:discussion}, we discuss the main findings of our work,
its relations to other work,
  and we suggest follow-up work.
\Sect{sec:conclusion} concludes the paper with an exposition of our most important findings.

\section{Theory and Methods}
\label{sec:Theory}

The present study addresses the collective response of $N$ tipping elements
that are coupled linearly through network links distributed according to
paradigmatic network types (Erdös-Rényi, Barabási-Albert, Watts-Strogatz).
The tipping elements are described by a system of differential equations adapted from Refs.~\citep{Kroenke2020, Wunderling2020} using the software package \textit{PyCascades}~\citep{wunderling2021modelling}.
A Gaussian white noise is applied to the tipping elements to model stochastic fluctuations.

\subsection{Equations of motion}
  
Many natural systems show tipping paired with hysteresis-like behavior~\citep{Brummitt2015}.
The behavior of such a tipping element is commonly modeled by cubic differential equations~\citep{Wunderling2020, Kroenke2020, Klose2020}
with normal form~\cite{Murdock2006}
\begin{align} \label{eq:normal_single_node}
  \frac{\rmd x}{\rmd t} = -x^3 + x + r 
\end{align}
where $r$ is the bifurcation parameter.
The system is in a bistable state when its right-hand side has three roots, i.e., for
$r \in (r_{\text{crit,low}}, r_{\text{crit,high}}) = (-\sqrt{4/27} , \sqrt{4/27} )$. 
At $r_{\text{crit,low}}$ and $r_{\text{crit,high}}$
two of the roots disappear in a saddle-node bifurcation.
The bifurcation diagram of a single node including noise is shown in Fig.~\ref{fig:model}a.

Commonly, tipping elements are not isolated.
For instance, \citet{Kroenke2020} suggested to study the dynamics of $N$ coupled tipping elements $k\in \{1,\dots ,N\}$ 
that are linearly coupled to other nodes $l\neq k$ 
(see Fig.~\ref{fig:model}b),
\begin{subequations}
\begin{align}
  \label{eq:system}
  \frac{\rmd x_k}{\rmd t} = - x_k^3 + x_k + {\sum_l}' d_{kl} x_l + \noiseStrength \frac{\rmd W_k}{\rmd t} \, .
\end{align}
The prime at the sum indicates that $l$ takes values in $l \in \{1,\dots ,N\} \backslash \{k\}$. On average a node $k$ is coupled to $pN$ other elements. $W_k$ denotes a Wiener process and $\noiseStrength$ the strength of the noise.

Equation~\eqref{eq:system} describes 
homogeneous systems where all
nodes follow the same dynamics and have the same critical parameter
values for tipping.
  Such systems are discussed in \Sect{sec:evolution} and \Sect{sec:late-time-evolution}.
In \Sect{sec:drift-impact} we 
expand that discussion to treat
  dynamics 
  with
additional drift terms, $d_k$,
and offset terms, $r_{kl}$,
\begin{align}
  \label{eq:drifted-system}
  \frac{\rmd x_k}{\rmd t}
  = - x_k^3 + x_k + d_k
  + {\sum_l}' d_{kl} \bigl( x_l - r_{kl} \bigr)
  + \noiseStrength \frac{\rmd W_k}{\rmd t} \, .
\end{align}
\end{subequations}
%

\subsection{Network topologies}

There is ample room
for different types of interaction networks
when one only specifies
that a node is connected to $p\,(N-1)$ other nodes on average.
We will therefore
consider three types of networks to explore the impact of the structure of the interaction network on the tipping dynamics:
\begin{description}
\item[ER]
  The connections form an Erdös-Rényi network~\citep{Erdos1959}
  when they are assigned randomly with probability~$p$.
  The probability distribution to be connected to $m$ nodes, i.e., the degree distribution,
  amounts to a binomial distribution. 

\item[BA]
  Barabási-Albert networks~\citep{Barabasi-Albert} are built by sequentially adding connections under the constraint of preferential attachment:
  elements with a higher number of connections have a larger chance to receive more links.
  This leads to scale-free networks
  where the degree distribution has a power-law tail that decays like $m^{-3}$.

\item[WS]
  Watts-Strogatz~\citep{Watts-Strogatz} networks are built by arranging the elements as a one-dimensional ring,
  where each node is connected to its $pN$ nearest neighbors.
  Subsequently, a fraction $\beta$ of the links is reconnected at random to a new element. 
  This generates networks with the small-world property. 
\end{description}

\subsection{Numerical Methods}

Tipping behavior in ER, BA, and WS networks has been analyzed previously by \citet{Kroenke2020}.
\citet{wunderling2021modelling} developed the Python package \textit{PyCascades} to
create the directed network of coupled tipping elements with the python \textit{networkx2.3} package~\citep{Hagberg2008} and analyze the dynamics of tipping cascades based on the package sdeint~\citep{sdeint} for the integration of stochastic differential equations.
These programs are used here to generate the networks and 
integrate the stochastic differential equations \eqref{eq:system}.
If not stated otherwise the system is integrated up to $t=100$, using a step size of $\Delta t=0.01$.

The coupling strengths $d_{kl}$ will be drawn from a uniform distribution with $0< d_{kl} < 2d$.

In the simulation runs,
the nodes are initially placed at $x=0$,
i.e., the unstable fixed point,
and the cascade is started by setting $\nu$ nodes to $x=1$. 
Thus, we focus on the transient behavior of tipping.

At times, the noise turns the point $x=0$ into a stable fixed point of the dynamics \citep{Arnold1992}.
Then, we study cases where the coupling of the system is strong enough
such that all nodes follow the initial displacement of the displaced node.

\section{Mean-Field Evolution}
\label{sec:evolution}

We characterize the state of the network of tipping elements by the
expectation
of the state $x_k$ of the elements,
\begin{subequations}
\begin{align}
  \label{eq:Xexpection}
  \xAv
  = \mean{x_k}
  = \frac{1}{N} \; \sum_{k=1}^N x_k 
\end{align}
and by their variance
\begin{align}
  \label{eq:Xvariance}
  \vAv
  = \mean{ (x_k - \xAv)^2 }
  = \mean{ y_k^2 }
\end{align}
\end{subequations}
where $y_k = x_k - \xAv$ denotes the deviation of $x_k$ from the expectation $\xAv$.
In \Sect{subsec:moment-expansion} we derive the equations for the time evolution of $x$ and $v$
that derive from \Eq{system}.
To this end we adopt a closure where we suppress higher-order moments and correlations.
Then we point out universal aspects of the time evolution towards the asymptotic steady state
(\Sect{subsec:order-parameter-asymptotics}).

\begin{figure*}
  \centering
  \includegraphics[width=0.9\textwidth]{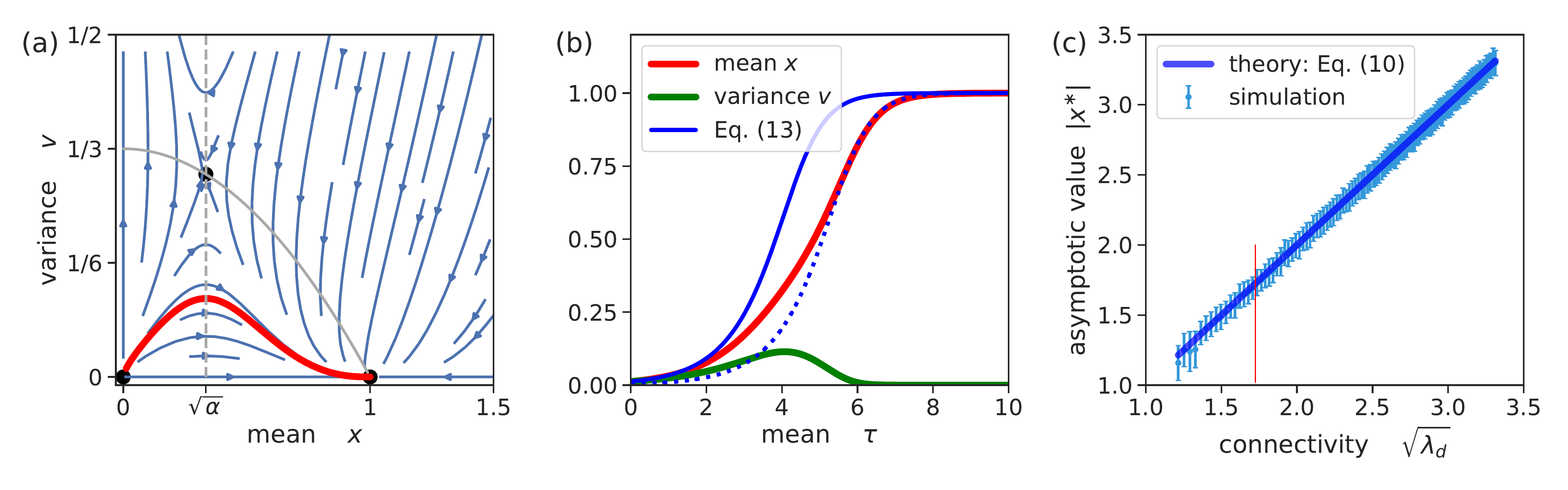}
  \caption[Absolute Asymptotic Values vs.~Connectivity $\connectivity$.]{
    \textit{Asymptotic Values of the Evolution.} 
    (a) Phase-space flow of the dynamical system, \Eq{evolution}, for $\sqrt\alpha \simeq 1/3$. The analytical solutions of the nullclines are marked by gray lines, and fixed points are marked by black circles, respectively. The red line shows the numerical solution for $N=80$ tipping elements that are connected by an ER network with $d=0.1$ and $p=0.25$. The numerical solutions for the blue vectors have been computed using the function \texttt{matplotlib.pyplot.streamplot} in python. (b) The red and the green line show the time evolution of the expectation and the variance of the numerical solution shown in panel~(a). The thinner blue line shows the prediction, \Eq{x-prediction}, for $x(0) = 1/N$ (for explanation see Sect.~\ref{subsec:x-approach-test}). The dotted line shows the same function shifted by $\Delta\tau = 1.24$. The shifted curve provides an excellent description of the late-time evolution of the expectation. (c) The data points show the asymptotic values $x^\ast$ of numerical simulations of \Eq{system} with $p=0.25$ and couplings $d_{kl}$ drawn from a uniform distribution $d_{kl}\in(0,0.2)$. Error bars indicate the standard deviation derived from $50$ realizations of the network configuration and the noise. The connectivity $\connectivity$ is varied by increasing $N$ from $20$ to $400$ (cf.~\Eq{connectivity}). The numerical values agree very well with the prediction \Eq{x*-settling} that is shown by a solid blue line. The parameter value of the trajectory shown in panels~(a) and (b) is indicated by a vertical red line. For small networks the system occasionally tips into the direction opposite to the perturbation. Hence, we plot the absolute value of $x^\ast$.
  }
  \label{fig:asymptotic_values}
\end{figure*}
\subsection{Moment expansion}
\label{subsec:moment-expansion}

In order to evaluate the time evolution of the
cumulants of the $x_k$ distribution 
we
note that  
taking averages~\mean{\_} is a linear operation
such that it commutes with taking the time derivative,
\begin{align*}
  \dot \xAv
  &= \frac{\rmd}{\rmd t} \mean{x_k}
    = \mean{ \dot x_k }
  \\
  \dot \vAv
  &= \frac{\rmd}{\rmd t} \mean{ y_k^2 }
   = 2 \: \mean{ y_k \: \dot x_k }.
\end{align*}
In order to evaluate the averages we express \Eq{system} as
\begin{align}  \nonumber
  \dot x_k
  &= - ( \xAv + y_k )^3 + \xAv + y_k 
    +  {\sum_l}' d_{kl} \bigl( \xAv + y_l \bigr)
    + \noiseStrength \frac{\rmd W_k}{\rmd t}
  \\ \nonumber
  &= \xAv \, \left( 1 - \xAv^2 + {\sum_l}' d_{kl} \right)
    + y_k \: ( 1 - 3\, \xAv^2 )
    - 3\, \xAv \, y_k^2 
    \\   \label{eq:xk-dot}
    &- y_k^3 +  {\sum_l}' d_{kl} \, y_l
    + \noiseStrength \frac{\rmd W_k}{\rmd t}.
\end{align}
Taking the average of \Eq{xk-dot}
and observing that $\mean{y_k}=0$
provides
\begin{align} \label{eq:dot-xAv}
  \dot \xAv
  &= \xAv \, \left( 1 - \xAv^2 - 3\, \vAv +  \mean{ {\sum_l}' d_{kl} } \right)
  \\ \nonumber
  & 
  - \mean{  y_k^3 }
  + \mean{ {\sum_l}' d_{kl} \, y_l }
  + \mean{ \noiseStrength \frac{\rmd W_k}{\rmd t} }.
\end{align}
The expressions in the second line of this equation account for
the third cumulant of the distribution of the state variables $x_k$,
a biased average of the deviations~$y_k$ from the average of the state variables,
and the impact of noise on the dynamics, respectively.
In the present study we consider a closure of the dynamics
where the third cumulant and the biased average will be suppressed. 
Moreover, we focus on the average behavior of an ensemble of systems, where we perform two averages at the same time, (i) we average over the coupling strength in the network realizations, (ii) and we average over the noise $W_k$.
Hence, the noise term vanishes,
and $\mean{ {\sum_l}' d_{kl} }$ takes the value $p \, (N-1)\, d$,
since each node is connected to $p \, (N-1)$ other nodes 
with an average coupling strength $d$.
Potential subtleties involved in taking the different averages will be addressed in a forthcoming paper. 
Altogether we thus find
\begin{subequations} \label{eq:moment-evolution}
\begin{align} 
  \dot \xAv
  &= \xAv \, \Bigl( 1 + p \, (N-1)\, d - \xAv^2 - 3\, \vAv  \Bigr).
\end{align}

The evolution of $v$ is obtained by multiplying \Eq{xk-dot} by $2 y_k$ and taking the average,
\begin{align*} 
  \dot \vAv
  = 2 \mean{ y_k \, \dot x_k }
  =& 2\, \vAv \,  ( 1 - 3\, \xAv^2 ) + \mean{ y_k \: \noiseStrength \frac{\rmd W_k}{\rmd t} }
  \\
  &- 6 \, \xAv \, \mean{ y_k^3 } - 2 \, \mean{ y_k^4 } + 2 \, \mean{ {\sum_l}' d_{kl} \, y_k \, y_l }.
\end{align*}
The term involving noise vanishes on account of considering a small-noise limit of the dynamics.
The terms in the second row of the equation are higher-order moments and correlations
that are dropped due to the assumptions of our closure of the moment hierarchy.
Hence, we obtain
\begin{align}
  \label{eq:v-evolution}
  \dot \vAv = 2\, \vAv \,  ( 1 - 3\, \xAv^2 ).
\end{align}
\end{subequations}

\subsection{Asymptotics of the order parameter}
\label{subsec:order-parameter-asymptotics}

The dynamics, \Eq{moment-evolution}, involves a single dimensionless parameter,
\begin{align} \label{eq:connectivity}
  \connectivity = 1 +  p \, (N-1)\, d \, ,
\end{align}
that depends on the average number of connections, $p\, (N-1)$, of a tipping element
and on the average coupling strength, $d$.
Hence, we denote it as \emph{connectivity.}

Rescaled variables
that are based on the connectivity 
will allow us to discuss the asymptotics of the order parameter in a more transparent form
\begin{align} 
  x(\tau) & = \xAv / \sqrt{\connectivity} \nonumber \\
  y(\tau) &= \vAv / \connectivity         \label{eq:nondimensionalization} \\
  \tau &= \connectivity \, t                 \nonumber
\end{align}
Denoting the 
derivatives of $x(\tau)$ and $y(\tau)$ with respect to $\tau$
as $\dot x$ and $\dot y$ 
provides
\begin{subequations}\label{eq:evolution}
\begin{align}\label{eq:x-evolution}
  \dot x
  &= \frac{\rmd x}{\rmd\tau}
   = x \, \bigl( 1 - x^2 - 3\, v \bigr)
  \\[1mm]
  \dot v
  &= \frac{\rmd v}{\rmd\tau}
   = 6\, v \, \bigl( \alpha - x^2 \bigr)
\end{align}
\end{subequations}
where $\alpha = (3\connectivity)^{-1}$.

In \Fig{asymptotic_values}(a) we show the phase-space flow of \Eq{evolution}.
The evolution has nullclines with $\dot x = 0$ at $x=0$ and $v=(1-x^2)/2$
that are marked by solid gray lines. 
The nullclines with $\dot v = 0$ lie at $v=0$ and $x=\sqrt\alpha$.
They are marked by dashed lines.
The three fixed points,
$p_0 = (0,0)$, $\left( \sqrt\alpha, (1-\alpha)/3 \right)$, and $(1,0)$,
are located at the intersections of
the $\dot v = 0$ and $\dot x = 0$ nullclines. 

By inspection of the flow crossing the nullclines one readily verifies
that the fixed point $p_i$ has $i$ stable directions.
For all $\alpha>0$ the trajectories
approach a state with
$x=\pm 1$ and vanishing variance. 
A representative sample trajectory is given by a solid red line in \Fig{asymptotic_values}(a).
Panel~(b) shows the time evolution of $x(\tau)$  (red line) and $v(\tau)$ (green line)
of this trajectory.

The approach towards the stable fixed point at $x=1$ implies that
\begin{align}  \label{eq:x*-settling}
  x^\ast
  = \lim_{\tau\to\infty} \xAv
  = \sqrt\connectivity
  = \sqrt{ 1 + d\, p\, (N-1) }
\end{align}
\Fig{asymptotic_values}(c) shows that this dependence is indeed observed by our numerical data.
Only for small connectivities, $\connectivity$, the data tend to lie closely below the prediction.
We will come back to this point
when we discuss Fig.~\ref{fig:tipping_time_shift}.

\section{Late-time evolution}
\label{sec:late-time-evolution}

The $\dot x = 0$ nullcline always takes the same form (solid gray line in \Fig{asymptotic_values}(a)), 
while the $\dot v = 0$ nullcline is a vertical (dashed) line whose position,
$x_c = \sqrt\alpha = (3\,\connectivity)^{-1/2}$,
depends on the system parameters. 
For trajectories that proceed to the left of $x_c$ 
the order parameter $x$ remains small while $v$ diverges.
This behavior is unphysical
and it lies out of the scope of our model
because our closure assumption can only be expected to work
for small~$v$. 
However, for large $N$ the fixed point $p_1$ lies very close to $(0,1/3)$,
and we only consider initial conditions where $0 < v < x \ll 1$.
For these initial conditions one encounters trajectories as shown by a solid red line.
Its $x$ coordinates grows monotonically from zero towards one.
Initially, the variance grows rapidly.
Subsequently, (after crossing the nullcline at $x = \sqrt\alpha$)
it decays towards zero.
For larger networks
\textcolor{black}{($N \gg 1$)}
and for more strongly coupled networks
(increasing $pd$)
the crossover arises 
very close to zero ($\sqrt\alpha = [3\, (N-1)\, p\,d]^{-1/2} \ll 1$),
and the maximum value of the variance decreases.

\subsection{Approach to the stable fixed point}
\label{subsec:fixed-point-approach}

The approach towards $p_2$ is governed by the linearized \Eq{evolution},
of the deviations $\epsilon = x-1$ and $v$ from the fixed point.
We take into account that $\alpha \simeq 1/3$ for large
connectivity, $\connectivity$,
and find 
\begin{align}   \label{eq:p2-stability}
  \begin{pmatrix} \dot\epsilon \\ \dot v \end{pmatrix}
  \simeq
  \begin{pmatrix} -2 & -3 \\
                   0 & -4
  \end{pmatrix}
  \begin{pmatrix} \epsilon \\ v \end{pmatrix} \, .
\end{align}
The eigenvalue for the decay in $v$ is therefore twice as large as the one in~$x$,
such that $v \sim (1-x)^2$ close to the fixed point;
in line with the flow lines of the phase-space plot
shown in \Fig{asymptotic_values}(a).

\begin{figure*}
  \centering
  \includegraphics[width=0.9\textwidth]{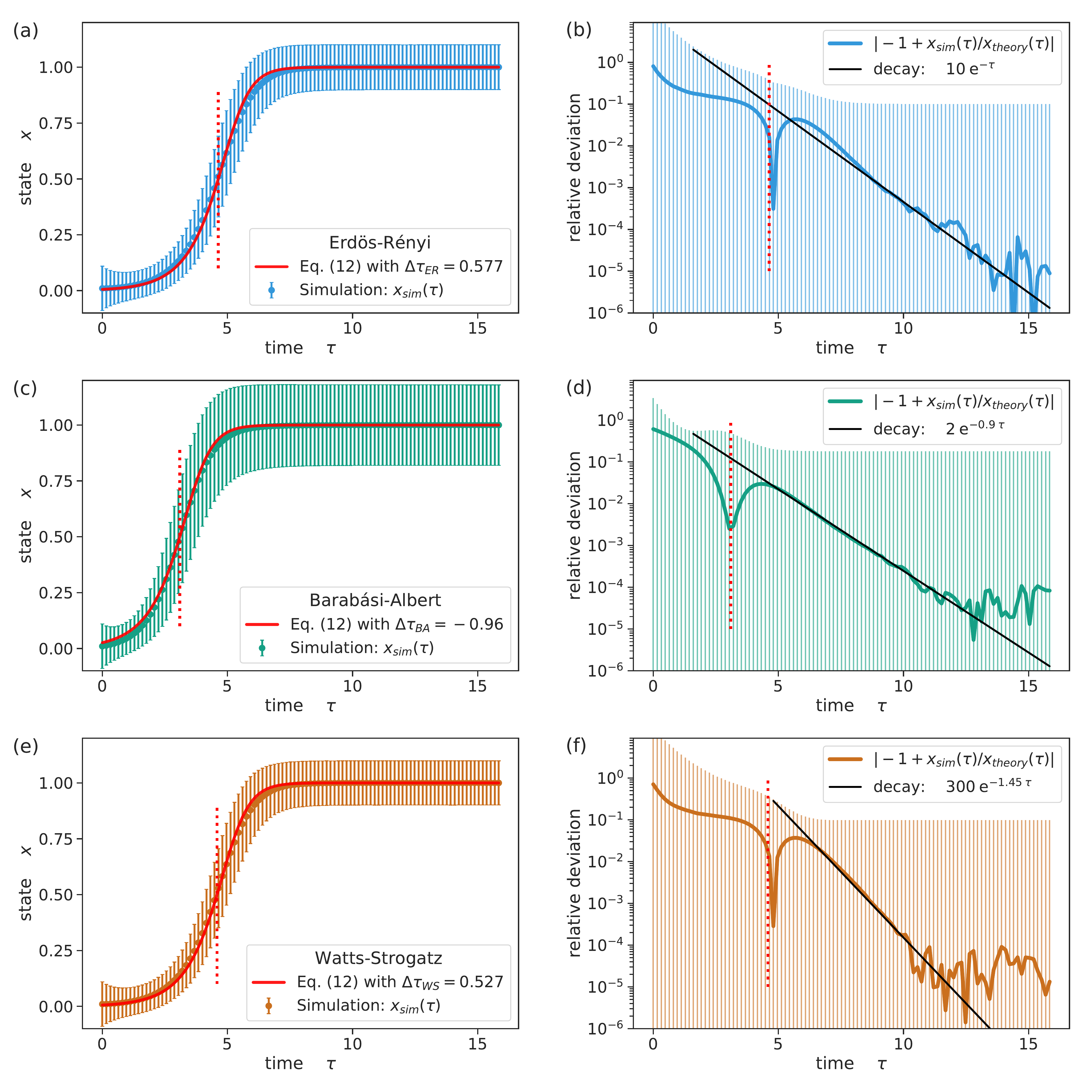}
  \caption[Collective Tipping Behavior of Different Network Types]{ 
    \textit{Collective Tipping Behavior of Different Network Types.}
    The left and right panels show the evolution of the order parameter
    and relative deviation of numerical data from \Eq{system} and the prediction from \Eq{x-prediction}.
    The rows show data for
    (a,b) Erdös-Rényi,
    (c,d) Barabási-Albert, and
    (e,f) Watts-Strogatz networks with reconnection probability $\beta=0.8$.
    The other parameters of the simulations are fixed to $N=100$, $p=0.1$, $d=1.5$, and $\noiseStrength = 0.1$,
    such that $\connectivity \simeq 16$. 
    The data points and the error bars represent the expectation and the 
    standard deviation over $50$ realizations of the network and the noise, respectively.
    The prediction, \Eq{x-prediction}, is indicated by a solid line in the left panels.
    It is shifted by $\Delta\tau$ to match the data at $x=0.5$ (see main text for details).}
  \label{fig:network_types}
\end{figure*}

In order to discuss the evolution of $x$ we introduce the variable $w=x^2$
and observe that
\begin{subequations} \label{eq:w-evolution}
\begin{align} 
  \dot w &= 2 \, w \: \bigl( 1 - w - 3\,v \bigr) \\
  \dot v &= 2 \, v \: \bigl( 1 - 3\, w \bigr) 
\end{align}
\end{subequations}
The evolution of $w(\tau)$ for identically vanishing variance, $v \equiv 0$, is readily obtained by variable separation and partial fraction decomposition
\begin{align*} 
  2 \, \tau
  &= \int_{\wI}^{w(\tau)} \frac{ \rmd w }{ w \: ( 1 - w ) }
    =  \int_{\wI}^{w(\tau)} \rmd w \: \left[ \frac{1}{w} + \frac{1}{1-w} \right]
  \\
  &= \ln\left[ \frac{w(\tau)}{\wI} \: \frac{ 1-\wI }{1-w(\tau)} \right]
\end{align*}
such that
\begin{align}   \label{eq:x-prediction}
  x(\tau) = \sqrt{ w(\tau) } =
  \left[ 1 + \frac{ 1-\xIsq }{\xIsq} \: \rme^{-2\, \tau} \right]^{-1/2} \, .
\end{align}
For late times, $\tau \gg 1$, we hence obtain the expected exponential decay
\begin{align}   \label{eq:x-asymptotics}
  1 - x(\tau) \simeq  \frac{ 1-\xIsq }{ 2\, \xIsq} \; \rme^{-2\, \tau}
  \qquad\text{ for \ } \tau \gg 1 \, .
\end{align}

In order to discuss the impact of finite values of $v$ 
we look for the solutions of \Eq{w-evolution} for $w=1-\varepsilon$
where
\begin{align*} 
  \dot\varepsilon &= -2 \, \varepsilon \, \bigl( 1 - \varepsilon \bigr) - 6\,v \, \bigl( 1 - \varepsilon \bigr)  \\
  \dot v          &= -4 \, v \: \left( 1 + \frac{3}{2}\,  \varepsilon \right) 
\end{align*}
We
approximate
the equation for $\dot\varepsilon$ by inserting
$v = c \, \varepsilon^2$ 
(cf.~the discussion of the relation of $v$ and $\epsilon$ below \Eq{p2-stability}),
and consider terms only till order $\varepsilon^3$ and $v\, \varepsilon$.
As a consequence
\begin{align*} 
  (1+3c)\, \dot\varepsilon &= -2 \, (1+3c)\, \varepsilon \, \bigl( 1 - (1+3c)\, \varepsilon \bigr) 
\end{align*}
such that $\varepsilon(\tau) = 1-x(\tau)$ remains unchanged up to multiplication by the factor $1+3c$. 
However, in view of \Eq{x-asymptotics} 
such a factor can be accounted for by a shift, $\Delta\tau$, of time, yielding the following prediction for the time evolution of the order parameter,
\begin{align}   \label{eq:xTheory}
  x_{\text{theory}} (\tau)
  \simeq 
  \left[ 1 + \frac{ 1-\xIsq }{\xIsq} \: \rme^{-2\, (\tau-\Delta\tau)} \right]^{-1/2} \, .
\end{align}
This expression has a single free parameter $\Delta\tau$
  that amounts to a time delay to reach the half time $\tau_{1/2}$,
  where $x_{\text{theory}} (\tau_{1/2}) = 1/2$,
  \begin{align}   \label{eq:tauHalf}
    \tau_{1/2}
    = \Delta\tau - \frac{1}{2} \: \ln\frac{3 \, \xIsq}{ 1-\xIsq } \, .
  \end{align}

\subsection{Numerical test of \Eq{xTheory}}
\label{subsec:x-approach-test}
The left columns of \Fig{network_types} show the evolution of the order parameter for ER, BA, and WS networks.
As initial condition we consider a system where one node,
let it be $\ell$, is displaced to a value $x_\ell=\sqrt{\connectivity}$,
and all other nodes, $m\neq \ell$, start at $x_m=0$.
This amounts to initial conditions with
\begin{align*} 
  \xI
  &= \frac{ \mean{ x_k }_{\tau=0} }{\sqrt\connectivity}
   = \frac{1}{\sqrt\connectivity} \: \frac{1}{N} \sum_{k=1}^N x_k(0) \\
  &= \frac{0\times (N-1) + 1\times \sqrt{\connectivity}}{N\,\sqrt\connectivity}
   = \frac{1}{N}  \, .
\end{align*}
For each type of network we perform $50$ runs with different realizations of the networks and noise.
The expectation of the order parameter, $x_{\text{sim}}$, for the resulting evolution is given by a circle,
and the error bars mark the standard deviation 
over the simulation runs. 

The theoretical prediction, $x_{\text{theory}}$, is provided by a solid line.
%
The time 
offsets are determined 
such that the numerical data and theoretical prediction match for $x=1/2$.
In a logarithmic plot of the relative deviation $(x_{\text{sim}}/x_{\text{theory}})/x_{\text{theory}}$ (\Fig{network_types}(b,d,f))
this results in a singularity at
time $\tau_{1/2}$
that is marked by vertical dotted lines.
The lines align with the singularity for
\begin{align}  
  \Delta\tau_{\text{ER}} &= 0.577 \, , \nonumber
  \\
  \Delta\tau_{\text{BA}} &= -0.96 \, , \label{eq:offset-times}
  \\
  \Delta\tau_{\text{WS}} &= 0.527 \, . \nonumber
\end{align}
ER and WS networks have 
very similar time offsets of about $\Delta\tau \simeq 0.5$.
In contrast BA networks decay significantly faster.
  The negative value of $\Delta\tau_{\text{BA}}$ implies that
these networks decay even faster than predicted by the $v=0$ result, \Eq{x-prediction}.
Follow-up work will have to clarify if and how this intriguing finding is related to the existence of hubs in BA networks,
nodes with a degree greatly exceeding the average.

In addition
to the position of $\tau_{1/2}$,
 \textcolor{black}{\Fig{network_types}(b,d,f)}
demonstrates
that the prediction, \Eq{xTheory}, provides an excellent description of the data for times beyond
$\tau_{1/2}$.
The time offset $\Delta\tau$ is the only fitting parameter in this description.

\subsection{Influence of Initial Displacement}
\label{subsec:initial_displacement}

\begin{figure*}
  \includegraphics[width=0.9\textwidth]{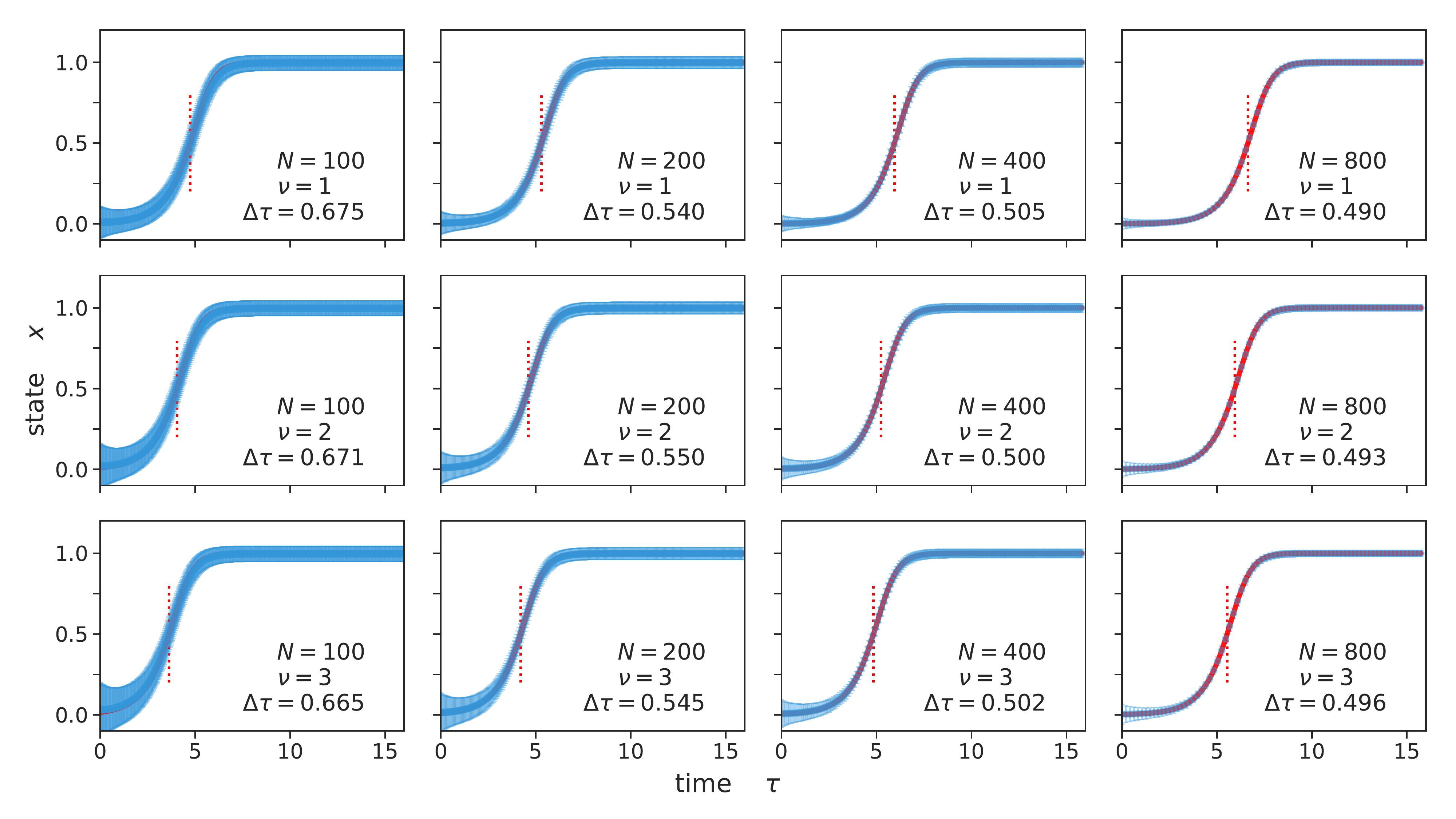}
  \includegraphics[width=0.9\textwidth]{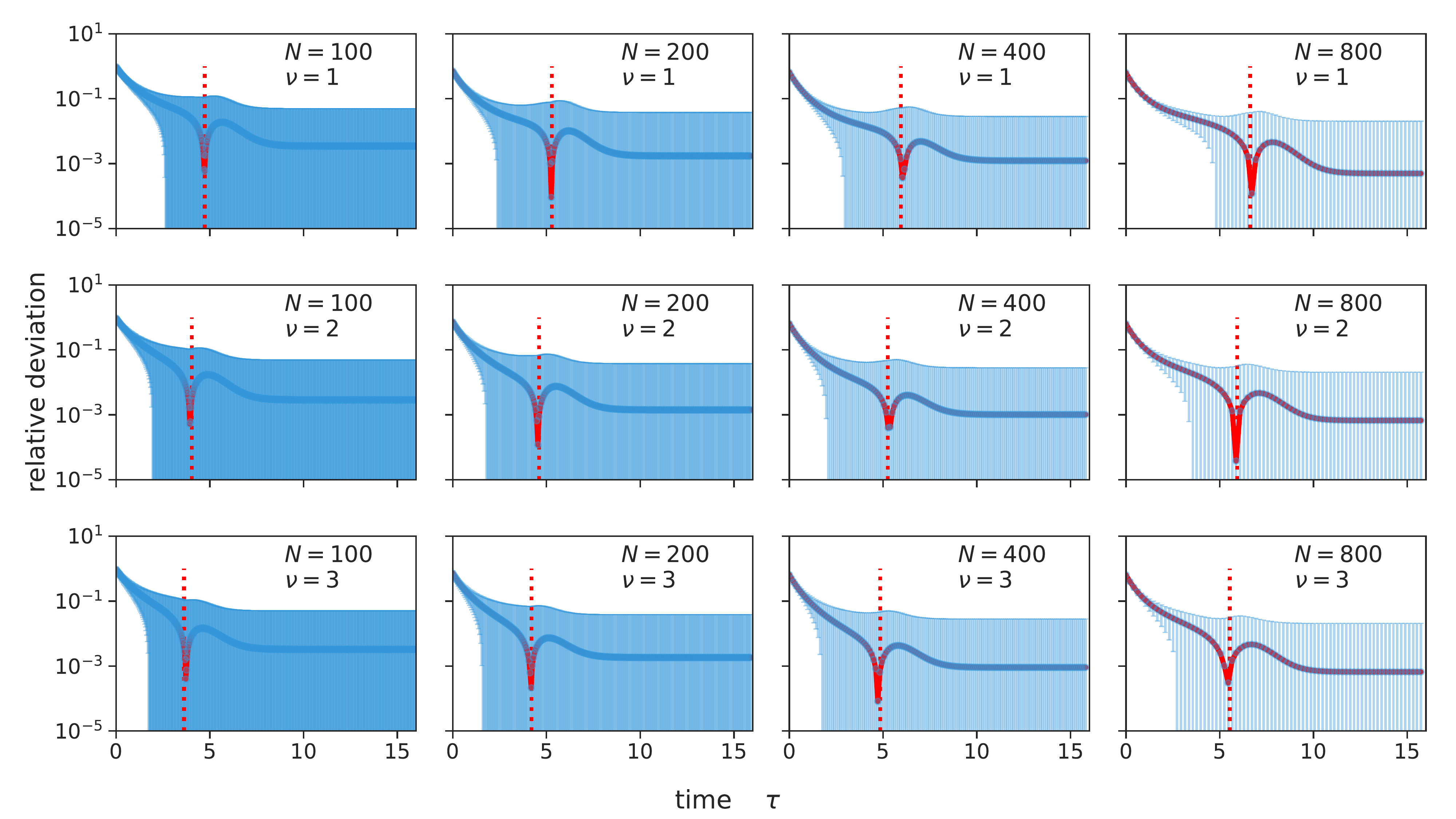}
  \caption{ 
  \textit{Trajectories for Different Initial Displacements $x(0) = {\nu}/{N}$.}
  Calculations using \Eq{system} have been conducted
  for ER networks with 
  $d=0.1$,~$p=0.25$,~$\noiseStrength=0.1$,
  $N\in \{100,200,300,800\}$,
  and~$\nu \in \{1,2,3\}$,
  with couplings $d_{kl}$ drawn from a uniform distribution $d_{kl}\in(0,0.2)$.
  This amounts to connectivities of $\connectivity = 3.475$, $5.975$, $10.975$, and $20.975$, respectively.
    The data and their error bars in the upper panels 
    represent the expectation and standard deviation of 
  $100$ simulations runs with independent realizations of the network and the noise.
  The corresponding lower panels shows the relative deviation of the trajectory from the  prediction
  \Eq{xTheory} with $x_0 = \nu/N$ and 
  a time shift $\Delta\tau$ 
  provided in the respective legends.}
\label{fig:tipping_time_shift}
\end{figure*}

In the previous sections one node has been perturbed by a fixed value initially.
To discuss the effect of
larger initial perturbations 
we explore 
now systems
where initially $\nu$ nodes are tipped to the asymptotic equilibrium value $x^\ast = \sqrt\connectivity$.
Consequently, 
\begin{align} \nonumber
  \xI
  &= \frac{ \mean{ x_k }_{t=0} }{\sqrt\connectivity}
    = \frac{1}{\sqrt\connectivity} \: \frac{1}{N} \sum_{k=1}^N x_k(0)
  \\ \label{eq:x-0-nu}
  &= \frac{0\times (N-\nu) + \nu\times \sqrt{\connectivity}}{N\,\sqrt\connectivity}
   = \frac{\nu}{N}  \, .
\end{align}
According to \Eq{xTheory} this change in $x(0)$
will be the only change in $x_{\text{theory}}$.
This 
prediction is scrutinized 
in Figure~\ref{fig:tipping_time_shift}
where we show data for ER networks with $100$, $200$, $400$, and $800$ nodes
  where we excited initially $\nu=1$, $2$ or $3$ nodes:
  \begin{enumerate}
  \item
    For a given value of $N$ the time offsets for different $\nu$
    \textcolor{black}{are} indeed constant to within our numerical accuracy.
    Moreover, the positions of the dotted lines clearly show the decrease of $\tau_{1/2}$
    that is predicted by \Eq{tauHalf}. 
  \item
    For increasing $N$ the time offset decreases.
    This is in line with the expectation that the variance reaches smaller values for increasing $\connectivity$
    such that the dynamics more closely follows the $v=0$ dynamics.
  \item
    The relative deviation does not decay to zero.
    Indeed on a second sight one recognizes that the data in \Fig{network_types}
    also tend to saturate at a $10^{-5}$ level.
    Hence, the data do not quite saturate at $\sqrt\connectivity$
    but at a slightly smaller value, as also observed for the small connectivity data in \Fig{asymptotic_values}.
    We attribute this effect to the small noise
    that enforces $v$ to take a small positive value.
    According to \Eq{x-evolution} this reduces the asymptotic value of $x$ by a factor of $\sqrt{1-3\,v^2}$.
    Consequently, the relative error decreases like $1-\sqrt{1-3v} \simeq 1.5 \vAv / \connectivity$.
    The trend is clearly visible in the data.
  \end{enumerate}
Altogether,
%
we established an analytic description for the time evolution of the order parameter 
of coupled networks subject to small noise levels.
For different paradigmatic network types,
the collective dynamics of the network behaves just as a single aggregated tipping element
with a new time scale and equilibrium values
selected by the connectivity $\connectivity$, as provided by the scaling in \Eq{nondimensionalization}.

\section{ Dynamics and stability beyond the normal form }
\label{sec:drift-impact}

Up to now we considered systems that evolve according to~\Eq{system}.
In this section we explore how the dynamics changes when there are additional drift and offset terms,
and how rescaling and shifting the variables $x_k$ affects the parameters of the resulting description. 

\subsection{Influence of  drift and offset terms }

To discuss the impact of drift and offset terms we write~\Eq{drifted-system} in the form of \Eq{xk-dot} with two additional terms,
\begin{align*} 
  \dot x_k
  =& \xAv \, \left( 1 - \xAv^2 + {\sum_l}' d_{kl} \right)
    + y_k \: ( 1 - 3\, \xAv^2 ) - 3\, \xAv \, y_k^2 
    \\
    & - y_k^3
      +  {\sum_l}' d_{kl} \, y_l
      + \noiseStrength \frac{\rmd W_k}{\rmd t}
  \\
  &
  + d_k - {\sum_l}' d_{kl} \, r_{kl} 
\end{align*}
The terms in the first row contributed to the average of~$\dot x_k$, as evaluated in \Eq{dot-xAv}.
The terms in the second row amount to higher-order cumulants, correlations, and noise
that are dropped for the closure and the weak-noise limit adopted here.
The terms in the third row account for the drift and offset terms. 
Let the expectation of the drift terms $d_k$ be $r_D$,
and for the offsets $r_{kl}$ in the coupling terms we introduce
\begin{align*} 
  \mean{ {\sum_l}' d_{kl} \, r_{kl} }
  = p\,(N-1) \, d \, r_O
  = (\connectivity -1) \, r_O \, .
\end{align*}
(Again, there may be correction terms that are subdominant for the considered closure.)
With these notations the average of $\dot x_k$ takes the form
\begin{align}   \label{eq:xAvDot}
  \dot \xAv
  &= \xAv \, \Bigl( \connectivity 
    - \xAv^2 - 3\, \vAv  \Bigr)
    + r_D - (\connectivity - 1) \, r_O
\end{align}
and $\dot\vAv$ still takes the form of \Eq{v-evolution}.
After all, the additional terms are constant
such that 
they average to zero when multiplied by $y_k$.
In terms of the rescaled variables, \Eq{nondimensionalization} we hence find
\begin{subequations}
\begin{align}  \label{eq:x-dynamics-biased}
  \dot x &= x \, \bigl( 1 - x^2 - 3\, v \bigr)
           + 
           \frac{r_D+r_O}{\connectivity^{3/2}} - \frac{r_O}{\sqrt\connectivity} \, ,  \\
   \dot v &= 6\, v \, \bigl( \alpha - x^2 \bigr) \, ,
\end{align}
\end{subequations}
We note that for large
connectivity $\connectivity = 1+p\,d\,(N-1)$
the drift and offset terms are sub-dominant in the order-parameter dynamics,
and they disappear in the large-network limit.
Hence, the order parameter of strongly coupled, large networks evolves according to \Eq{evolution},
as 
discussed in \Sect{sec:evolution}.

For large connectivity, $\connectivity \gg 1$, every system
that crosses a tipping point relaxes to its new equilibrium value according to these equations. 
The intrinsic order-parameter dynamics is symmetric. 
In terms of the effective coupling
\begin{align}  \label{eq:effCouping}
  r_{\text{eff}} =  \frac{r_D - r_O}{\connectivity^{3/2}} - \frac{r_O}{\sqrt\connectivity}
\end{align}
one 
encounters saddle-node bifurcations at $r_{\text{eff}} = \pm \sqrt{4/27}$
(cf.~\Fig{model} and the discussion below \Eq{normal_single_node}).

\subsection{ Linear coordinate transformations }
\label{ssec:linTrafo}

Scale changes
can be expressed as a change of variables where
the state of the nodes is characterized by 
new variables $z_k$
that 
are 
linear functions of~$x_k$,
\begin{align}   \label{eq:def-z}
  z_k = m \, x_k + n
  \qquad\text{for fixed \ } m, \: n \, .
\end{align}
Consequently, 
$\zAv = \mean{ z_k } = m \, \xAv + n$
and $\dot z_k = m \, \dot x_k$.
Averaging the latter equation and using \Eq{xAvDot} provides 
\begin{align*} 
  \frac{\dot\zAv}{m}
  = \dot\xAv
  = \xAv \: \left( \connectivity - \xAv^2 - 3\, \vAv \right) + r_D - (\connectivity-1) \: r_O
\end{align*}
For
$z = \zAv / (m\sqrt\connectivity)$,
$\tau = \connectivity\, t$, and
$z_0 = n / (m\sqrt\connectivity)$ we have $x = z-z_0$,
and 
we find
\begin{subequations}
\begin{align}   \label{eq:z-dynamics}
  \dot z(\tau) 
  &= \frac{\rmd z}{\rmd\tau }
  = (z-z_0) \: \left[ 1 - (z-z_0)^2 -3\, v \right] +  r_{\text{eff}} 
  \\
  \dot v(\tau)
  &= 6\, v \: \left( \alpha - (z-z_0)^2 \right)
\end{align}
\end{subequations}
with a macroscopic bifurcation parameter,
$r_{\text{eff}}$, given by \Eq{effCouping}.

The shift $z_0$ of $z$ can be interpreted as a vertical displacement of the s-shaped red curve in \Fig{model}.
It affects the positions of the steady states, but not the bifurcation parameters. 
The intrinsic dynamics remains symmetric in the bifurcation parameter $r_{\text{eff}}$
with saddle-node bifurcations at $r_{\text{crit}} = \pm \sqrt{4/27}$.

The dependence of the bifurcation parameter, \Eq{effCouping},
on connectivity $\connectivity$ provides another important insight into the stability of the networks:
The drift dominates the control parameter as long as the offset terms $r_{kl}$ are very small as compared to the drift,
$ (\connectivity - 1 ) \: |r_O| \ll |r_D| $, 
and the global system is bistable for $-\connectivity^{3/2} \lesssim 2.5\, r_D \lesssim \connectivity^{3/2}$
(after all $\sqrt{27/4} \approx 5/2$).
In view of \Eq{connectivity} the connectivity $\connectivity$
increases linearly 
with the network size,
such that the state of the network in this parameter range is very robust against perturbations.
However, for strong connectivity the offset terms will always dominate the system response eventually,
\begin{align*} 
  (\connectivity - 1 ) \: |r_O| \gg |r_D|
  \quad \text{for \ } \connectivity \ggg 1 \, . 
\end{align*}
Eventually,  
the coupling terms govern the strength of the bifurcation parameter,
and the range of stability increases 
more slowly with the connectivity
\begin{align*} 
  -\sqrt{ {\textstyle\frac{4}{27}} \: \connectivity } \; \lesssim \: r_O \: \lesssim \; \sqrt{ {\textstyle\frac{4}{27}} \: \connectivity }
  \quad \text{for \ } \connectivity \ggg 1 \, . 
\end{align*}
Still, the range of stability increases with network size and
for increasing network size 
the state 
will
become ever more robust against perturbations.

\begin{figure*}
  \includegraphics[width=0.9\textwidth]{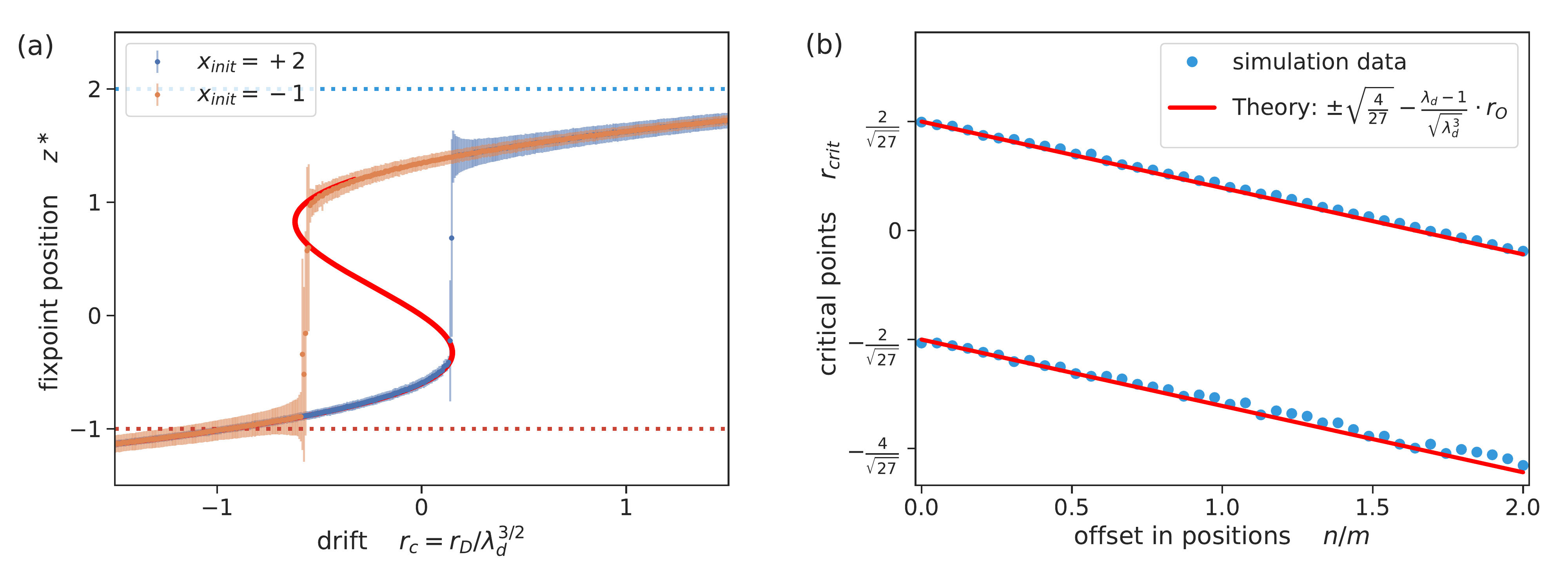}
  \caption[Critical Points for Different $\xI$]{ 
    \textit{Asymptotic fixed point $z^\ast$ (a) and the critical points (b) for networks with an offset $n$ and $r_{kl}=0$ in \Eq{climate-system}.}
    All calculations are conducted for an ER network with $N=100$ nodes, couplings $d_{kl}$ drawn from a uniform distribution $d_{kl}\in(0, 1.2)$, $\connectivity=16$, and noise strength $\noiseStrength=0.1$. (a) For fixed values $n=m=1$ and different drift $r_D$ we follow the dynamics for initial conditions $x_{\text{init}}=-1$ and one excited node at $z=1$ (orange) and $x_{\text{init}}=2$ and one node at $z=-1$ (blue), respectively. The dots and the error bars indicate the average and the standard deviation of the late-time asymptotic positions of fifty simulations with different realizations of the network and noise. They lie nicely on the prediction for the positions of the fixed points, \Eq{fixPtShift}, that is provided by the solid red line. The positions of the critical points, $r_{\text{crit}}$ are identified as having the largest distance between subsequent data points. (b) The $n$-dependence of  $r_{\text{crit}}$ determined from data as shown in panel (a), $m=1$, and different values of $n$. The red lines provide the predicted positions, \Eq{bifuPt}, of the saddle-node bifurcations.}
  
\label{fig:x0_relation}
\end{figure*}
\subsection{ Coordinate transformation that do not involve the coupling }

It is by no means self-evident
that a change of coordinates of individual tipping elements
affects the coupling terms in the same way. 
In particular, 
studies of climate networks \citep{Wunderling2021interacting} 
address the situation
where the individual tipping elements are shifted at a \emph{fixed} symmetric coupling.
In other words one adopts \Eq{def-z} but considers the dynamics
\begin{align} \nonumber 
  \frac{\rmd}{\rmd t} \frac{z_k}{m} 
  =& - \left( \frac{ z_k - n }{m}  \right)^3 + \frac{ z_k - n }{m} + d_k
  \\ \label{eq:climate-system}
  &+ {\sum_l}' d_{kl} \bigl( z_l - r_{kl} \bigr)
  + \noiseStrength \frac{\rmd W_k}{\rmd t} 
\end{align}
with $\mean{ r_{kl} } = 0$.
To identify the effects of this asymmetry in the parameter change we observe that
\begin{align*} 
  {\sum_l}' d_{kl} \bigl( z_l - r_{kl} \bigr)
  = {\sum_l}' (m\, d_{kl})\: \left( \frac{z_l-n}{m} - \frac{r_{kl} - n }{m} \right)
\end{align*}
such that the substitution $d_{kl} \to m \, d_{kl}$ and $r_{kl} \to (r_{kl}-n)/m$
provides the system treated in \Sect{ssec:linTrafo}.
On the level of the coarse-grained system this entails that for the present system
$\connectivity = 1 + (N-1) \, p \, md$
and
$r_O = \mean{(r_{kl}-n)/m } = -n/m = - \sqrt{\connectivity} \, z_0$.
Substitution into \Eq{z-dynamics} provides
\begin{align*} 
  \dot z
  &= (z-z_0) \: \left[ 1 - (z-z_0)^2 -3\, v \right] + \frac{r_D + r_O}{\connectivity^{3/2}} - \frac{r_O}{\sqrt\connectivity}
  \\
  &= (z-z_0) \: \left[ 1 - (z-z_0)^2 -3\, v \right] + \frac{ \connectivity-1}{\connectivity} \: z_0 + r_c
\end{align*}
with control parameter
\begin{align*} 
  r_c = {r_D}/{\connectivity^{3/2}} \, .
\end{align*}
Now, the fixed points, $\dot z = 0$, are located on the line
\begin{subequations}
\begin{align}   \label{eq:fixPtShift}
  r_c =  \frac{r_D}{\connectivity^{3/2}}
  = -(z^\ast-z_0) \: \left[ 1 - (z^\ast-z_0)^2  \right] - \frac{\connectivity-1}{\connectivity} \: z_0 \, ,
\end{align}
and the saddle-node bifurcations arise when
\begin{align}   \label{eq:bifuPt}
    r_{\text{crit}} = \pm\sqrt{\frac{4}{27}} - \frac{ \connectivity-1}{\connectivity} \: z_0 \, .
\end{align}
\end{subequations}
Figure~\ref{fig:x0_relation} demonstrates
that this is indeed observed in simulations of the networks.
The right-hand side of \Eq{fixPtShift} mounts to the red s-shaped curve in \Fig{x0_relation}(a). 
The data points mark the positions of the fixed points $z^\ast$
adopted by the system.
For each value of $z_0$ we thus find
a hysteresis loop that provides
the critical driving parameters, $r_{\text{crit}}$,
where the saddle-node bifurcations arise.
Figure~\ref{fig:x0_relation}(b) displays the $n$-dependence of 
the positions of the saddle-node bifurcations. 
The data nicely follow the expected behavior, \Eq{bifuPt},
with $z_0 = n/(m\sqrt\connectivity)$.
We conclude that incongruent response of the tipping elements and their coupling to parameter changes,
as expressed by \Eq{climate-system} and \eqref{eq:def-z},
gives rise to dynamics where the system is bistable in a range of control parameters $r_c$ that is no longer symmetric around zero. 

\section{Discussion and Outlook}
\label{sec:discussion}

In the main part of the present paper we set up an analytical description of the collective tipping of a large number, $N$, of tipping elements
that are coupled in a random network. 
The present section discusses these findings in the context of related literature,
and it highlights prospective extensions.

\subsection{Collective behavior}

We characterized each tipping element $k \in \{1\cdots N\}$ by a scalar state variable $x_k \in \mathbb R$.
It evolves in a double-well potential with a saddle at $x=0$
that separates the domains of attraction of two coexisting stable states at $x_k = \pm 1$.
The elements are pairwise connected in a random network
with mean coupling strength $d$, and a probability $p$
that there is a connection for a given pair of elements. 

The overall state of the system is characterized by the expectation \xAv\ and the variance \vAv\ of the distribution of $x_k$.
The evolution of these collective coordinates is derived based on a closure
where we drop higher-order cumulants and correlations.
In the weak-noise limit they evolve according to \Eq{moment-evolution}
that involves a single dimensionless parameter, the connectivity $\connectivity = 1 + p\, (N-1) \, d$. 
The coarse-grained dynamics has global fixed points at $(\xAv, \vAv) = (\pm\sqrt\connectivity, 0)$,
and tipping proceeds in terms of a dimensionless time $\connectivity \, t$
(\Fig{asymptotic_values}).

The dependence on $\connectivity\, t$ implies
that tipping occurs more abruptly for increasing network degree $p\, (N-1)$ and coupling strength $d$.
This provides a rational for the observation of \citet{Eom2018}
who also observed these trends in numerical data. 

The parameter dependence of the time scale and the position of the fixed points of the collective coordinates
can be absorbed by rescaling, \Eq{nondimensionalization}. 
In line with  the observations of \citet{Wunderling2020} this implies
that systems with high average degree $p\, (N-1)$ only require low coupling strengths $d$ to initiate a tipping cascade.

\textcolor{black}{We have analyzed network-specific time delays in \Eq{offset-times} and found very similar offsets for ER and WS networks of about $\Delta \tau \simeq 0.5$. Interestingly, BA networks decay with a significantly smaller time offset. We expect this to result from the existence of hubs in the BA network that could force the tipping process of the network. Follow-up work will have to clarify if and how this finding is related to the existence of hubs in the BA network. }

\subsection{Shape of the distribution}

In terms of the rescaled collective coordinates $x = \xAv / \sqrt\connectivity$ and $v = \vAv / \connectivity$ 
the dynamics has a saddle point at position $\Bigl(\alpha,(1-\alpha^2)/3 \Bigr)$ with $\alpha = (3\connectivity)^{-1/2}$.
Trajectories that proceed to the right of the saddle show collective behavior.
Up to initial transients the variance of the distribution of \textcolor{black}{trajectory states} is small
and the systems evolves according to a strongly-coupled collective dynamics.
Due to the assumptions taken to close the equations for $x$ and $v$ the present theory does not apply 
for trajectories that proceed to the left of the saddle.
Follow-up work will address higher-order closures
in order to deal with the broad and, at times, bimodal distributions of tipping elements in this polydisperse parameter regime.

At this point we observe that $\alpha$ approaches zero for increasing connectivity.
Hence, a rapidly increasing fraction of initial conditions proceed to the right of the saddle. 
In line with the findings of \citet{Brummitt2015}
this explains 
that increasing coupling strength $d$ entails synchronous behavior of tipping elements for a larger set of initial conditions.

\subsection{Noise}

Careful inspection of our data in \Fig{tipping_time_shift} revealed
that the expectation  \xAv\  saturates at values slightly below  $\sqrt\connectivity$.
We argued that the variance \vAv\ of the distribution remains finite for a system subjected to small noise,
and that this will decrease the asymptotic value by a factor $1-3\,\vAv/\connectivity$.
Moreover, preliminary data indicate
that noise has a non-trivial impact on the dynamics in the polydisperse parameter regime.
An expansion of our model
that \textcolor{black}{addresses the impact} of noise is in preparation.

\subsection{Negative feedback and multiple scales}
\label{ssec:negFeedback}

The derivation of the collective dynamics, \Eqs{moment-evolution},
does not involve any assumptions on the distribution of the coupling strength.
The present study describes the time evolution of systems
  where the feedback is positive on average, 
  $d>0$.
In this case \connectivity\ increases for larger and more strongly coupled networks,
and
the dynamics does not much depend on the structure and realization of the network
(cf.~\Fig{network_types}).

For negative overall feedback, $d<0$, and larger networks \textcolor{black}{our connectivity as defined in \Eq{connectivity} becomes negative and thus no longer meaningful. The nondimensionalization introduced in \Eq{nondimensionalization} to dimensionless parameters only works for $\connectivity>0$.
In the case of negative overall feedback,} the flow described by \Eqs{moment-evolution} will always enter the realm of large \vAv\
that is out of the scope of the present assumptions for the closure of the equations.
An appropriate expansion of the model will allow us to address phenomena such as Kadyrov-oscillations~\cite{kadyrov1984mathematical,abraham1991computational,wunderling2020basin}.

We also expect that systems with noticeable negative feedback
and vast heterogeneity in the time scales of the tipping elements and their tipping thresholds
are much more susceptible to correlations of the characteristics of tipping elements and their network environment.
Expansions that account for these correlations will make it possible to investigate systems like the
large-scale, Amazon rainforest,
where spatially distant patches are dependent on each other via the atmospheric moisture recycling feedback~\citep[e.g.][]{Wunderling2020,staal2018forest,Zemp2017}.
Instead of a micro-scale (local-scale tipping elements) and a macro-scale (entire forest),
one will then also define an intermediary meso-scale for strongly connected subcomponents of the forest.

\subsection{Time dependence of parameters}

The present analysis provides a comprehensive framework for the description of phase separation of networks of tipping elements
for fixed-in-time control parameters.
However, the Earth climate system is subjected to a sustained change of its parameters due to the release of vast amounts of carbon dioxide into the atmosphere,
and on historical scales its energy input varies periodically due to Milankovitch cycles \cite{HaysImbrieShackleton1976,SorensenEtAl2020}.
Both effects can have a severe impact on the time evolution.
Parameter oscillations in a noisy bistable system induce stochastic resonance \citep{BenziParisiSuteraVulpiani1982,GammaitoniHanggiJungMarchesoni1998}.
Parameter drift induces size focusing~\citep{Vollmer2014}
as discussed recently for 
applications in chemistry~\citep{ClarkKumarOwenChan2011} and soft biological matter~\citep{RosowskiSai-ZwickerStyleDufresne2020}.
Non-trivial new behavior emerges in the latter systems due to overall constraints on the dynamics
due to mass conservation, or global pressure and elastic fields.
It will therefore be highly relevant to explore how the dynamics is impacted by the coupling to a global parameter,
like temperature in the  Daisyworld models~\citep{Watson1983}.

\section{Conclusion}
\label{sec:conclusion}

In the present paper we established a description of the collective tipping dynamics of an assembly of coupled tipping elements
that are set up in a slightly perturbed unstable steady state. 
We described the assembly in terms of the expectation and the variance of variables
that describe the state of the individual tipping elements.
For a vast range of initial conditions of the network the variance remains small throughout the evolution.
In those cases the expectation follows universal dynamics,
as shown in Figs.~\ref{fig:network_types} and~\ref{fig:tipping_time_shift} for networks
with vastly different numbers of nodes, $N$, 
network degrees, $p\, (N-1)$,
network types (Erdös-Rényi~\citep{Erdos1959}, Barabási-Albert~\citep{Barabasi-Albert}, and Watts-Strogatz~\citep{Watts-Strogatz}),
and average coupling, $d$, between the elements.
The dynamics is characterized by a single dimensionless parameter,
$\connectivity$, that is defined in \Eq{connectivity}.
Based on the nondimensionalization, \Eq{nondimensionalization},
it provides the time scale of the dynamics
and the parameter values of the expectation and the variance of the target state of the dynamics
(\Fig{asymptotic_values}).
In \Sect{sec:drift-impact} we showed that the description is robust in the large network and strong coupling limit
(see \Fig{x0_relation}),
as long as the coupling provides a positive feedback on average.
Expected changes for negative feedback, $d<0$, and other extensions of the model were briefly discussed in \Sect{sec:discussion}.

In conclusion, in the present work we established
an effective analysis of 
cascading tipping behavior in strongly coupled networks.
It provides a comprehensive analytical description that is in excellent agreement with numerical simulations,
and it calls for extensions to address the tipping dynamics in other parameter regimes and in settings with a global feedback or other constraints.

\begin{acknowledgments}
  
  N.W. acknowledges support from the the IRTG 1740/TRP 2015/50122-0 funded by DFG and FAPESP.
  N.W. is grateful for a scholarship from the Studienstiftung des deutschen Volkes.
  J.F.D. is thankful for support by the Leibniz Association (project DominoES) and the European Research Council project Earth Resilience in the Anthropocene (743080 ERA).
  The authors gratefully acknowledge the European Regional Development Fund (ERDF), the German Federal Ministry of Education and Research and the Land Brandenburg for supporting this project by providing resources on the high performance computer system at the Potsdam Institute for Climate Impact Research.

\end{acknowledgments}

\bibliography{lit}

\end{document}